\documentclass[twocolumn,showpacs,preprintnumbers,amsmath,amssymb, superscriptaddress]{revtex4-1}

\usepackage{graphicx}
\usepackage{dcolumn}
\usepackage{amsmath,mathrsfs,amssymb,amsthm}
\usepackage{hhline}
\usepackage{wasysym,enumerate,color}
\usepackage{epstopdf}
\usepackage{verbatim}
\usepackage{hyperref}
\usepackage{color}
\usepackage{ulem} 

\renewcommand{\emph}[1]{\textit{#1}} 

\definecolor{darkgreen}{rgb}{0,0.5,0}
\definecolor{purple}{rgb}{0.35,0,0.35}
\definecolor{orange}{rgb}{1,0.5,0}
\definecolor{darkred}{rgb}{.7,0,0}
\definecolor{darkblue}{rgb}{0,0,.3}
\definecolor{grey}{rgb}{.6,.6,.6}
\definecolor{dimgreen}{rgb}{0.2,0.6,0.1}
\hypersetup{colorlinks,linkcolor=blue,urlcolor=blue,citecolor=blue}

\newcommand{\be}{\begin{equation}}
\newcommand{\ee}{\end{equation}}
\newcommand{\bea}{\begin{eqnarray}}
\newcommand{\eea}{\end{eqnarray}}

\newcommand{\bk}{{\bf k}}

\newcommand{\hn}{{\hat n}}

\newcommand{\cS}{{\cal S}}

\newcommand{\cE}{{\cal E}}
\newcommand{\cO}{{\cal O}}

\newcommand{\bra}[1]{\langle #1|}
\newcommand{\ket}[1]{|#1\rangle}

\newcommand{\e}{\varepsilon}

\newcommand{\s}{\sigma}

\begin{document}
\title{Fingerprints of Majorana fermions in spin-resolved subgap spectroscopy }
\author{Razvan Chirla}
\email{chirlarazvan@yahoo.com}
\affiliation{Department of Physics, University of Oradea, 410087, Oradea, Romania}
\author{C\u at\u alin Pa\c scu Moca}
\affiliation{Department of Physics, University of Oradea, 410087, Oradea, Romania}
\affiliation{BME-MTA Exotic Quantum Phase Group, Institute of Physics, Budapest University of Technology and Economics,
H-1521 Budapest, Hungary}

\date{\today}
\begin{abstract}

When a strongly correlated quantum dot is tunnel-coupled to a superconductor, it leads to the formation of Shiba bound states inside the superconducting gap. They  have been measured experimentally in a superconductor-quantum dot-normal lead setup. Side coupling the quantum dot to a topological superconducting wire that supports Majorana bound states at its ends, drastically affects the structure of the Shiba states and induces supplementary in-gap states. The anomalous coupling between the Majorana bound states and the quantum dot gives rise to a characteristic imbalance in the spin resolved spectral functions for the dot operators. These are  clear fingerprints for the existence of Majorana fermions and they can be detected experimentally in transport measurements. In terms of methods employed, we have used analytical approaches combined with the numerical renormalization group approach.
\end{abstract}


\maketitle

\section{Introduction}
Majorana bound states (MBSs) \cite{Alicea.11, Alicea.12} are zero energy states that appear at the ends of a topological superconductor with broken spin degeneracy.
During the last few years, there have been many proposals on how to probe the
MBSs. Signatures of MBSs have been found in ferromagnetic atomic chains assembled on the surface of superconducting lead (Pb)~\cite{Nadj-Perge.14}, in one dimensional semiconducting wires~\cite{Das.12}, in hybrid semiconductor-superconductor structures~\cite{Rokhinson.12, Mourik.12}, or at the vortex core inside a topological insulator superconductor heterostructure~\cite{Xu.15}. One of their distinctive features is the zero bias anomaly in tunneling experiments~\cite{Law.09, Sau.10}. On the other hand, this is also a hallmark for other phenomena, such as the Kondo effect in quantum dots (QD)~\cite{Goldhaber-Gordon.98}, the '0.7 anomaly' in a point contact~\cite{Rejec.06}, and the formation of the Shiba bound states in a superconductor-QD-normal metal (S-QD-N) 
setup~\cite{DeFranceschi.10, Deacon.10}. It is present even in systems 
where the Kondo effect and superconductivity compete~\cite{Buitelaar.02}. 
Therefore, finding ways for the clear confirmation of MBSs is a difficult task. 

When the same experimental response can be triggered by different phenomena, one promising route to differentiate between them is to study systems that can capture them both. Such an interplay between the in-gap Shiba states and the  MBSs has been successfully investigated in Ref.~\cite{Nadj-Perge.14}. 

\begin{figure}[!tb]
\includegraphics[width=0.9\columnwidth]{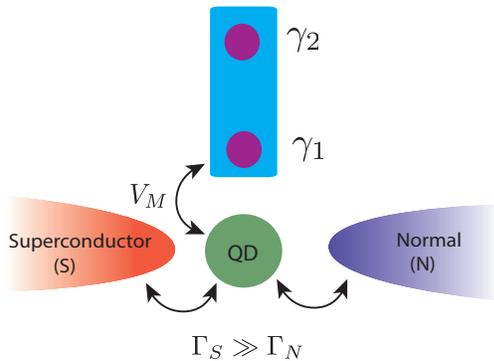}
\caption{(Color online)  Sketch with the setup. An interacting quantum dot (QD) is 
embedded between a superconductor and a normal lead and is side-coupled to a a topological 
superconducting wire that supports two Majorana modes $\gamma_{1}$ and $\gamma_{2}$ at the ends. The QD is strongly coupled to the superconductor, and weakly coupled to the normal lead, i.e., $\Gamma_{S}\gg \Gamma_{N}$. 
}
\label{fig:sketch}
\end{figure}

In the present work, following this route, we propose and
study theoretically a tunable setup in which the Shiba bound states can hybridize with the MBSs. The device is presented in Fig.~\ref{fig:sketch} and consists of a QD side coupled to a topological superconducting wire (TSW) that supports MBSs near its ends. The QD is embeded between a BCS superconductor and a normal lead, forming a S-QD-N setup~\cite{Deacon.10}. 
It is assumed that the QD is weakly coupled to the normal lead (no Kondo correlations 
are formed on the normal side as the corresponding Kondo temperature $T_{K}^{(N)}$ is 
exponentially suppressed), and strongly coupled to the superconductor 
($\Gamma_{S}\gg \Gamma_{N}$). In the absence of the TSW, this system has been thoroughly investigated by now, both experimentally~\cite{Deacon.10, Lee.12b, Lee.13b, Kumar.14, Zitko.15} and 
theoretically~\cite{Shiba.68, Rozhkov.99, Meng.09, Rodero.12, Zonda.15}. It presents two distinctive phases, i.e., a doublet and a singlet~\cite{Matsuura.77} separated by a quantum phase transition (QPT). Furthermore,
the evolution of the subgap Shiba states as function of the gate voltage was
successfully mapped by conductance measurements, and the agreement with the 
theoretical predictions is outstanding~\cite{Deacon.10}.

We have found that if a TSW is attached to the QD, as in Fig.~\ref{fig:sketch}, the in-gap spectrum gets substantially modified by the presence of the MBSs.
Moreover, by investigating the spectral functions for the QD operators, we have found a 
strong characteristic imbalance in the spin resolved spectrum in the vicinity of the QPT, 
which in principle can be measured experimentally by performing spin
polarized tunneling measurements. 
Our findings indicate that such a device may provide clear fingerprints for the presence of the MBSs. In terms of the methods we used, our analytical estimates have been supplemented by state of the art numerical renormalization group (NRG) calculations \cite{Wilson.75, Krishnamurthy.80}. 
\section{The Model} \label{sec:model}

We consider a QD that is coupled to a TSW. The QD itself is described
by a single spinful interacting level with energy $\varepsilon$  and Coulomb
repulsion $U$:
\begin{equation}
H_{\rm D} = \e \sum_\s d_\s^\dagger d_\s + U {\hat n}_\uparrow {\hat n}_{\downarrow}\,.\label{eq:H_D}
\end{equation}
Here $d_\s^\dagger$ is the creation operator for an electron with spin $\s$ in the dot
and $\hat n_{\sigma} $ is the occupation operator in the spin-$\s$ sector. 
Although the TSW is a complicated
mesoscopic object, it has been shown in Ref.~\cite{Alicea.11, Tijerina.15}  that it is possible
to construct an effective model that captures the essential physics of the 
QD-TSW by representing the wire in terms of its Majorana end states
\begin{eqnarray}
H_{M}&=& i\, \e_M \gamma_1 \gamma_2=\e_M(1+f^{\dagger}f) \,,
\end{eqnarray} 
where $\gamma_1=1/\sqrt{2}(f+f^{\dagger})$, $\gamma_2=i/\sqrt{2}(f-f^{\dagger})$ are 
the operators for the MBSs at the ends of the wire, and $f$ and $f^{\dagger}$ are some 
regular fermionic operators associated with the MBSs. The former satisfy the anticommutation relations $\{\gamma_1, \gamma_2\}=0$, while $\gamma_{1}^{2}=\gamma_{2}^{2}=1$. In our configuration we consider the mode $\gamma_{2}$ to reside at the far end of the wire such that only $\gamma_{1}$ hybridizes with the states in the dot (see Fig.~\ref{fig:sketch}) and may even leak into the dot \cite{Vernek.14}. When the TSW is in the 
topological phase, we assume that due the orientation of the Zeeman field in the TSW \cite{Alicea.11, Tijerina.15}, only the spin-down channel of the dot is coupled to $\gamma_{1}$ 
\begin{eqnarray}\label{eq:HDM}
H_{D-M} &=& V_M(d^\dagger_\downarrow \gamma_1 + \gamma_1 d_\downarrow) \, ,
\end{eqnarray}
with $V_{M}$ the tunneling amplitude between the QD and MBS.
The effective model that describes the QD-TSW is then given by
\be
H_{\rm eff} = H_{\rm D}+H_{\rm M}+H_{\rm D-M}.\label{eq:H_eff}
\ee
To capture the interplay between the MBSs and the Shiba states, 
the QD is embedded between a 
BCS superconductor on one side and a normal lead on the other side. In such a 
S-QD-N setup, the Shiba states~\cite{Shiba.68} are 
well resolved in the local spectral functions as long as $\Gamma_{S}\gg \Gamma_{N}$, 
with $\Gamma_{N(S)}/\hbar$ the superconducting (normal) tunneling rates. If $\Gamma_{N}\ll \Delta$, the Kondo temperature $T_{K}^{N}$ (due to the 
coupling to the normal lead) is vanishingly small, and the S-QD-N setup is qualitatively analogous with a S-QD system, with the normal lead acting simply as a probe (such as the tip of a STM) \cite{Deacon.10}. 
The superconducting lead is described by the BCS Hamiltonian
\begin{eqnarray} \label{eq:H_S}
H_{\rm S} &=& \sum_{\bk, \s} \xi_{\bk\s}c^\dagger_{ \bk \s}c_{ \bk \s}-(\Delta\, c^\dagger_{ \bk \uparrow} c^\dagger_{ -\bk \downarrow}+\mathrm{H.c.})\,.
\end{eqnarray}
The first term describes free fermions with dispersion $\xi_{\bk\s}$ and the second one
the superconducting correlations with the superconducting gap $\Delta$. The conduction band ranges from -D to D and the  density of states is considered to be constant, 
$\rho_{0}=1/2D$. Within our numerical calculations, $D$ is considered to be the energy unit. 

The coupling of the dot to the superconducting lead is described by the Hamiltonian
\begin{eqnarray} \label{eq:H_tun}
H_{\rm tun} &=& \sum_{\mathbf{k},\sigma} V (d^\dag_\s c_{\mathbf{k}\sigma} + c^\dagger_{\mathbf{k}\sigma} d_\s)\,,
\end{eqnarray}
where the tunneling rate between the QD and the normal superconductor $\Gamma=\Gamma_{S}=\pi\rho_0 V^2$.
Eqs.~\eqref{eq:H_eff},~\eqref{eq:H_S} and \eqref{eq:H_tun} 
define our model Hamiltonian
\begin{equation}\label{eq:H}
H=H_{\rm eff}+H_{\rm S}+H_{\rm tun}
\end{equation}
In what follows, we neglect the direct coupling between the Majorana lead with the 
normal superconductor, as well as the coupling between the QD and the p-wave continuum in 
the TSW.

\section{Superconducting atomic limit, $\Delta\to \infty$}
\begin{figure}[!tb]
\includegraphics[width=0.49\columnwidth]{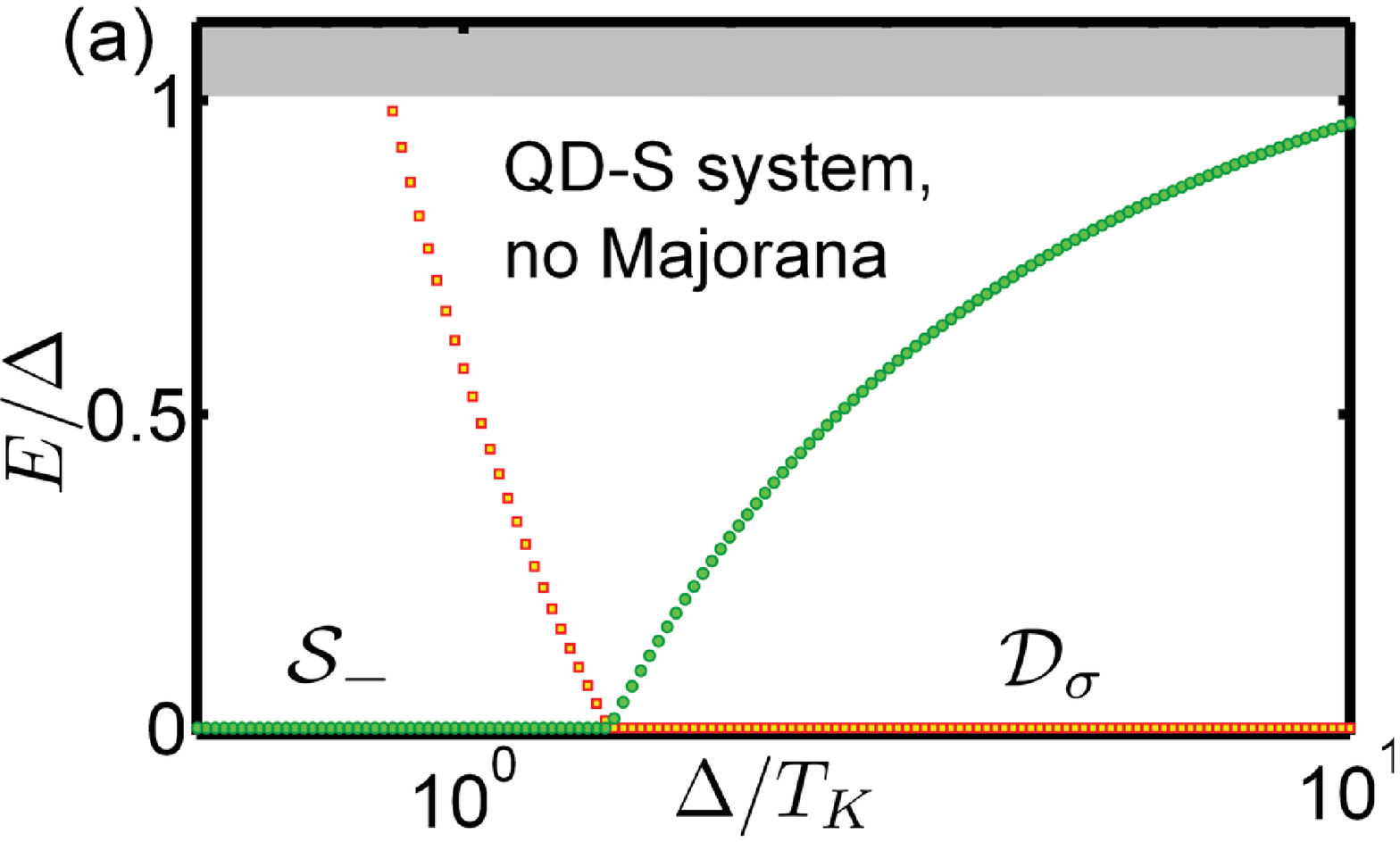}
\includegraphics[width=0.49\columnwidth]{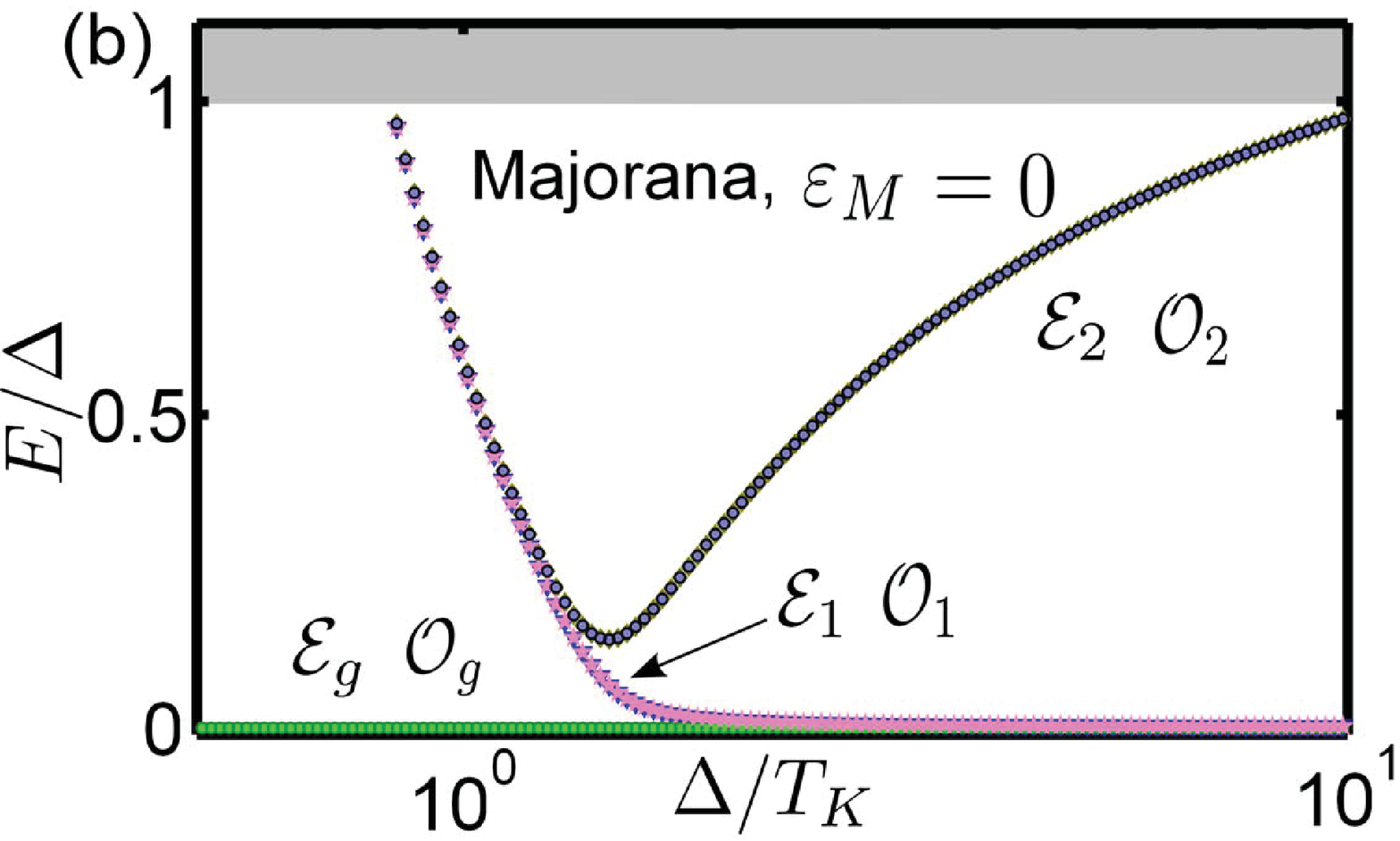}
\includegraphics[width=0.49\columnwidth]{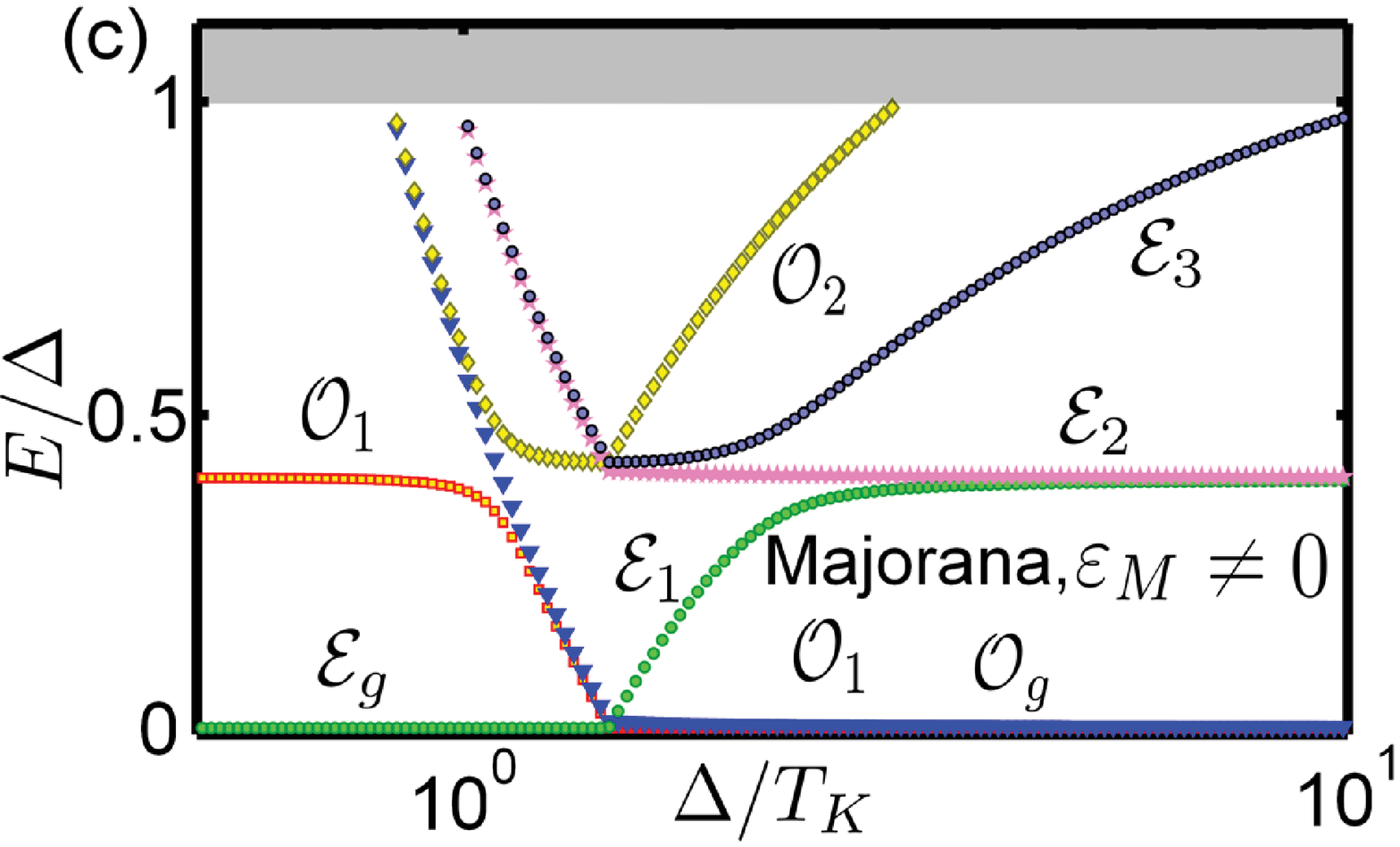}
\includegraphics[width=0.49\columnwidth]{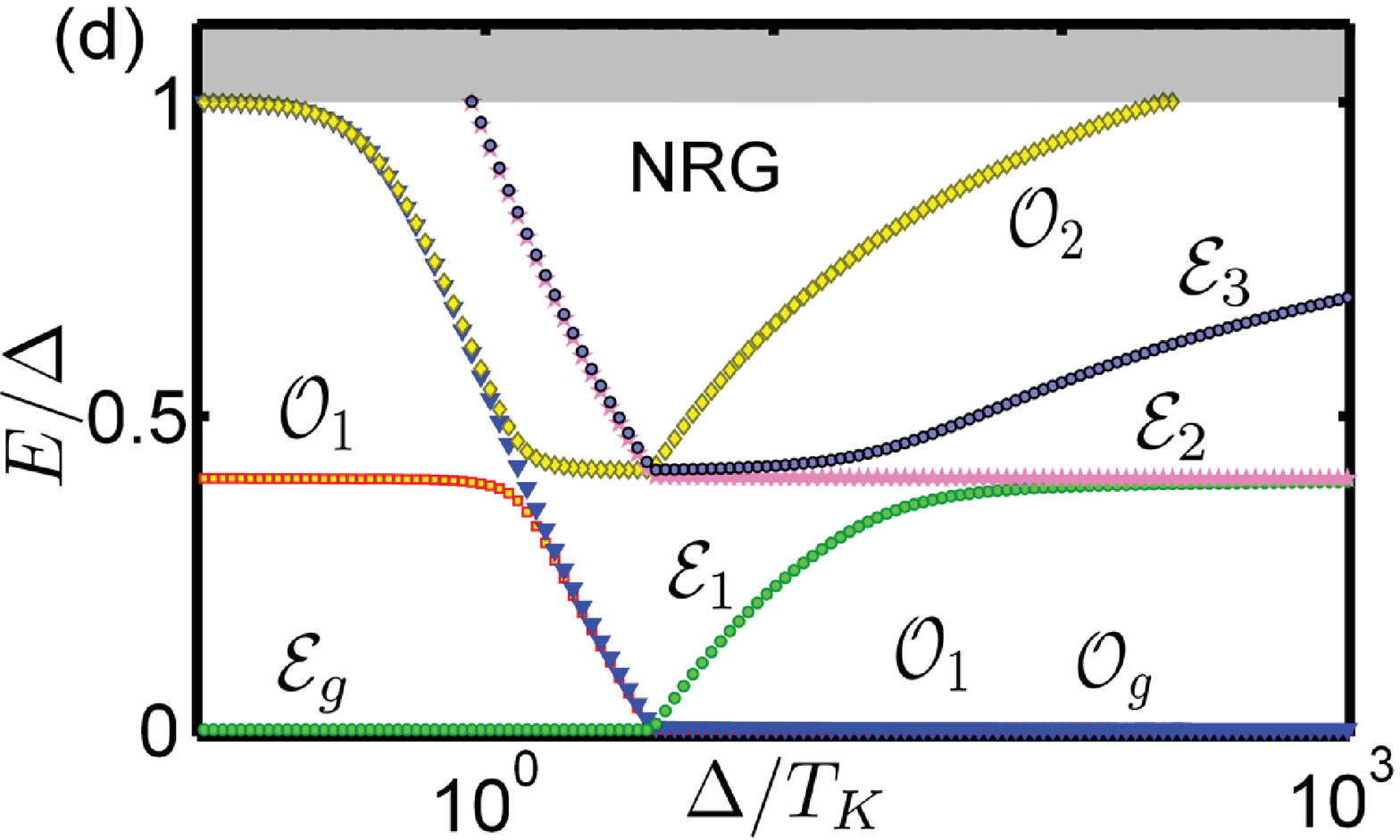}

\caption{(Color online) Analytical and NRG results for the  
energy spectrum in different configurations. (a) Bound states in the atomic limit for the S-QD setup, without an attached TSW.
The energies are those corresponding to Eq.~\eqref{eq:en_S_D1}. 
(b) The energy spectrum when the TSW is coupled to the QD and
the Majorana level has an energy $\e_M=0$. As the 
ground state has no particular symmetry, the QPT is absent.
(c) The evolution of the energy spectrum  when $\e_M= 0.4\,\Delta$. The level $\cO_{1}$ becomes degenerate with $\cO_{g}$
in the $\Delta/T_K\to \infty $ limit. 
(d) Similar to (c) but computed exactly with the NRG. Notice that we represent the 
evolution of the bound states on  a wider $\Delta/T_{K}$ range. 
In all the plots, the parameters  are: $V_M/\Delta=0.1$ and $U/\Delta=-2\e/\Delta=3$.
The shaded region above $\Delta$ represents the continuum.} 
\label{fig:levels_at}
\end{figure}

The model Hamiltonian introduced in Eq.~\eqref{eq:H} can not be solved exactly in general.
In the absence of the TSW, the Hamiltonian reduces to $H=H_{\rm D}+H_{\rm S}+H_{\rm tun}$, and the problem can be solved exactly by using the numerical renormalization group approach~\cite{Wilson.75, Krishnamurthy.80}. Analytically, the limiting case $\Delta\to\infty$, i.e., the superconducting atomic limit, has been addressed in several studies ~\cite{Meng.09, Bauer.07, Oguri.04, Tanaka.07}. By integrating the 
superconducting lead we can construct an effective Hamiltonian
that captures the essential physics. When the TSW is present, the Hamiltonian reads
\begin{multline}
H_{\rm el}^{(\Delta\to \infty)} = \sum_{\s}\xi_{d}\,d^\dagger_\s d_\s-\Gamma_{\varphi} \left (d^\dagger_\uparrow d^\dagger_\downarrow +\mathrm{H.c.}\right)+\\
+{U\over 2} \left (\hn-1 \right )^2+H_M+H_{D-M}.\label{eq:H_at_lim}
\end{multline}
Here,  $\xi_d= \e +U/2$ and 
$\Gamma_{\varphi}=2\,\Gamma/\pi \arctan\left (D/\Delta \right )$. The superconducting 
correlations are embedded in the second term in Eq.~\eqref{eq:H_at_lim}, which 
corresponds to a local pairing term induced by superconductivity in the dot. In what follows we shall restrict ourselves to the symmetric case $\e=-U/2$, but our main results also hold 
in the general case. Although more elaborate analytical models can be 
constructed~\cite{Zonda.15}, this simple model captures qualitatively all the important 
features, as demonstrated by a comparison with NRG calculations. 
The Hamiltonian~\eqref{eq:H_at_lim} can be diagonalized exactly and has in general
eight non-degenerate eigenstates.

To get a clear picture, let us first discuss 
what happens in the absence of the TSW. In this case, the system has a global $SU(2)\times Z_{2}$ symmetry corresponding 
to the conservation of the total spin in the dot (the Majorana modes have no spin index) and to the parity of the state \footnote{At half filling, when $\varepsilon=-U/2$, the system presents an extra $Z_{2}$ electron-hole symmetry}. This allows us to organize the states in spin multiplets. There are only four 
eigenstates that can be grouped into a pair of singlets $\ket {{\cal S}_{\pm}}$
and a doublet $\ket {{\cal D}_{\s}}$:
\begin{subequations}
\begin{align}
\ket {{\cal S}_\pm} &= 1/\sqrt{2}( \pm d^\dagger_\uparrow\, d^\dagger_\downarrow+1)\ket 0 \,,\label{eq:ket1}\\ 
\ket {{\cal D}_{\s}}&= d^{\dagger}_{\s} \ket {0}\,.\label{eq:ket2}
\end{align}
\end{subequations}
The corresponding eigenenergies are 
\begin{eqnarray}\label{eq:en_S_D1}
E_{\pm} &=& U/2 \pm \Gamma_{\varphi}\,,\;\;\;  E_\s = 0 \,.  
\end{eqnarray}
One notices that $E_{+}$ is always larger than $E_{-}$, and that the crossing point 
corresponding to $E_{-}=E_{\sigma}$  
signals a quantum phase transition (QPT), as the parity of the ground state changes. 
Although the QPT is captured in this simple local model, the true nature of the QPT is due to the competition between the Kondo screening and the superconducting correlations and happens at $\Delta\sim T_{K}$, with $T_{K}$ the Kondo scale~\footnote{We define $T_{K}$ by using Haldane's expression, $T_{K}=\sqrt{U\Gamma/2}\;\exp\;({-\pi U/(8\Gamma)})$}. 
Strictly speaking, the atomic limit is exact when $\Delta\to \infty$, but 
comparisons with the exact NRG results indicate that the superconducting atomic limit is a
good approximation as long as $\Delta$ is the largest energy scale in the problem. 
In Fig.~\ref{fig:levels_at}(a) we represent the evolution of the Shiba states in the
absence of the TSW as function of $\Delta/T_{K}$. When $\Delta\ll T_{K}$ the ground state
is the singlet  $\ket {\cal S_{-}}$, while in the opposite limit, $\Delta\gg T_{K}$, the 
ground state changes to the doublet $\ket {\cal D_{\sigma}}$. Such subgap resonances have 
been already measured in transport experiments in 
Refs.~\cite{Deacon.10, DeFranceschi.10} with 
good agreement with the theoretical predictions.

\begin{figure}[!tb]
\includegraphics[width=0.49\columnwidth]{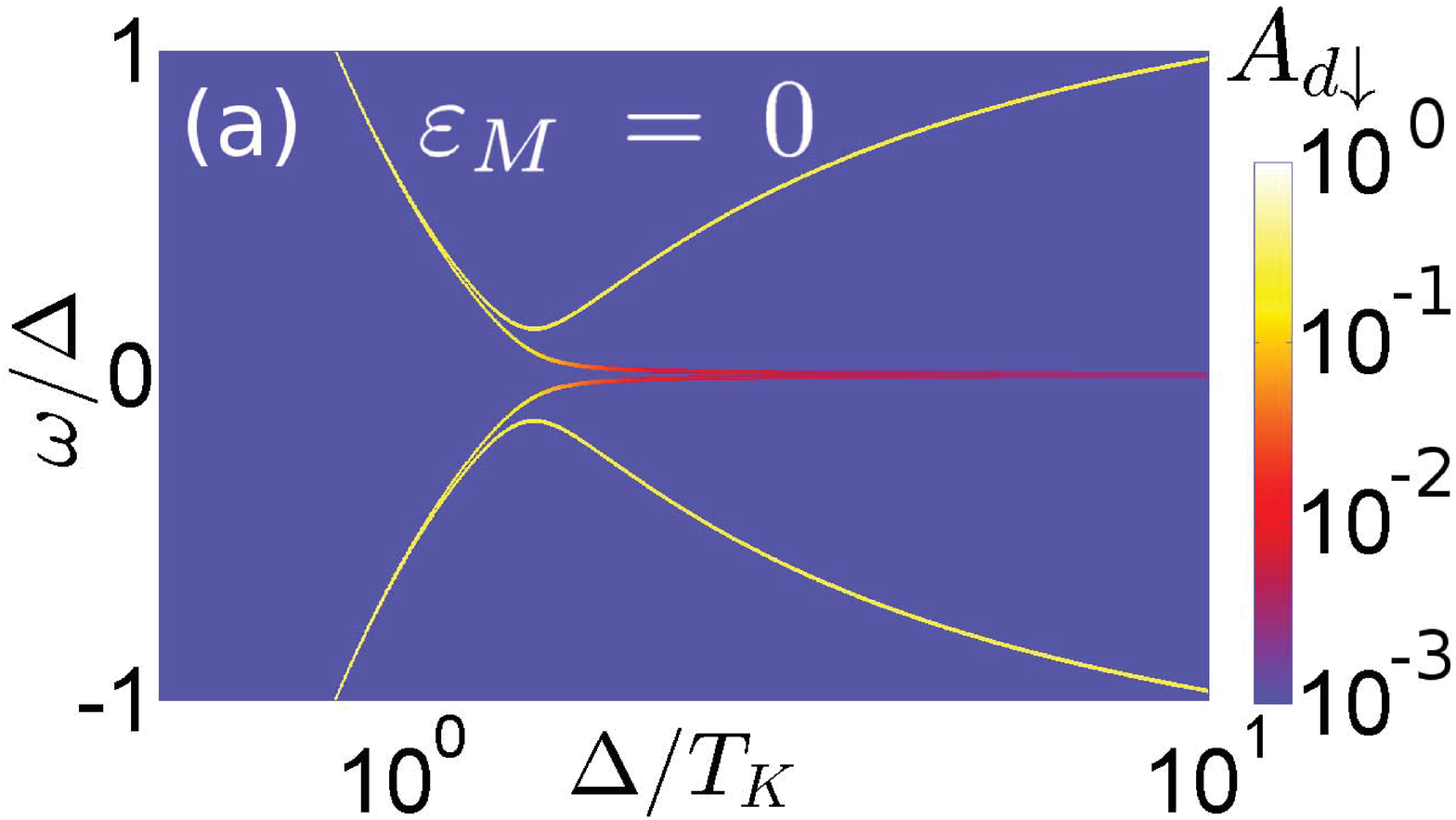}
\includegraphics[width=0.49\columnwidth]{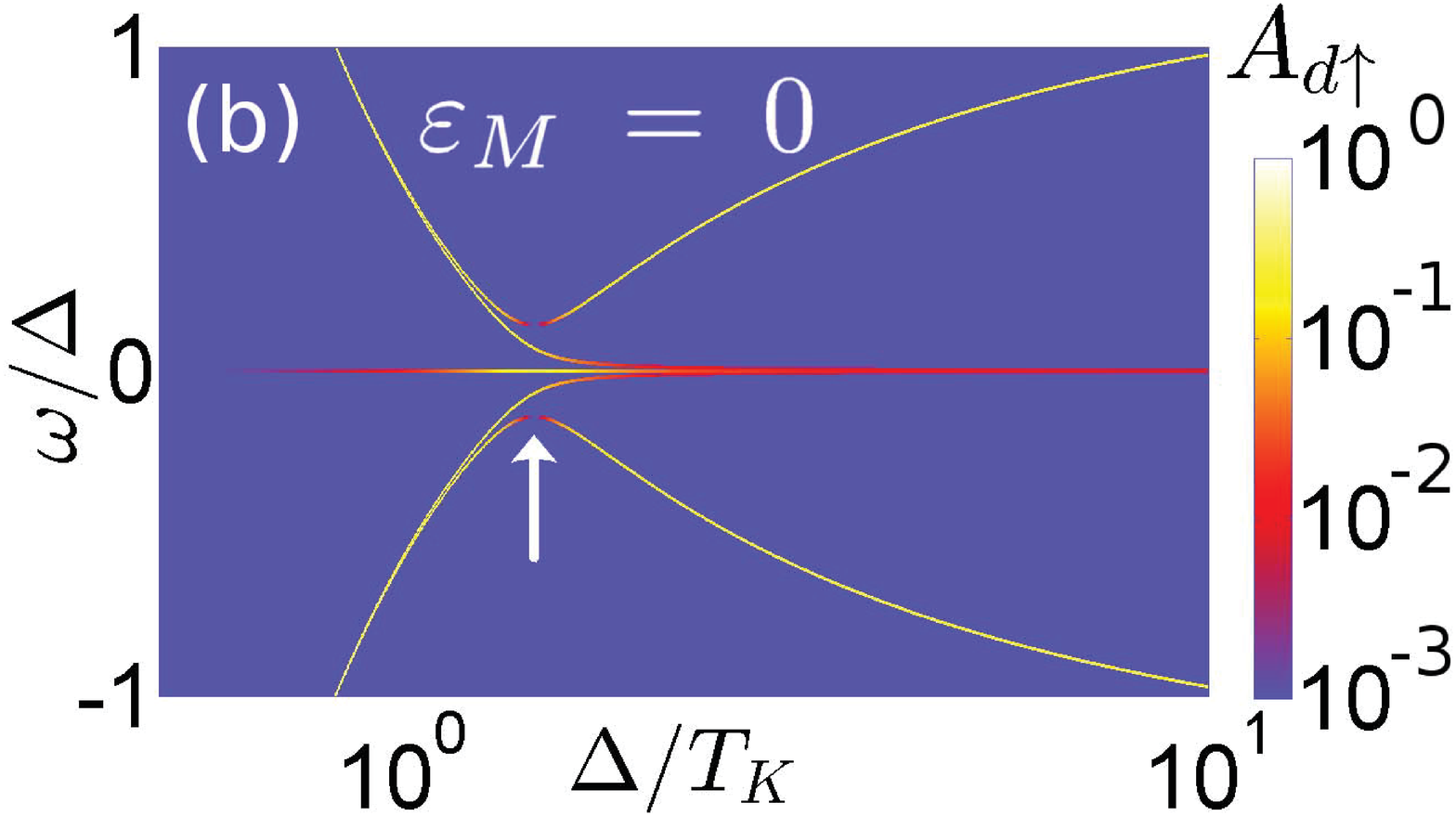}
\includegraphics[width=0.49\columnwidth]{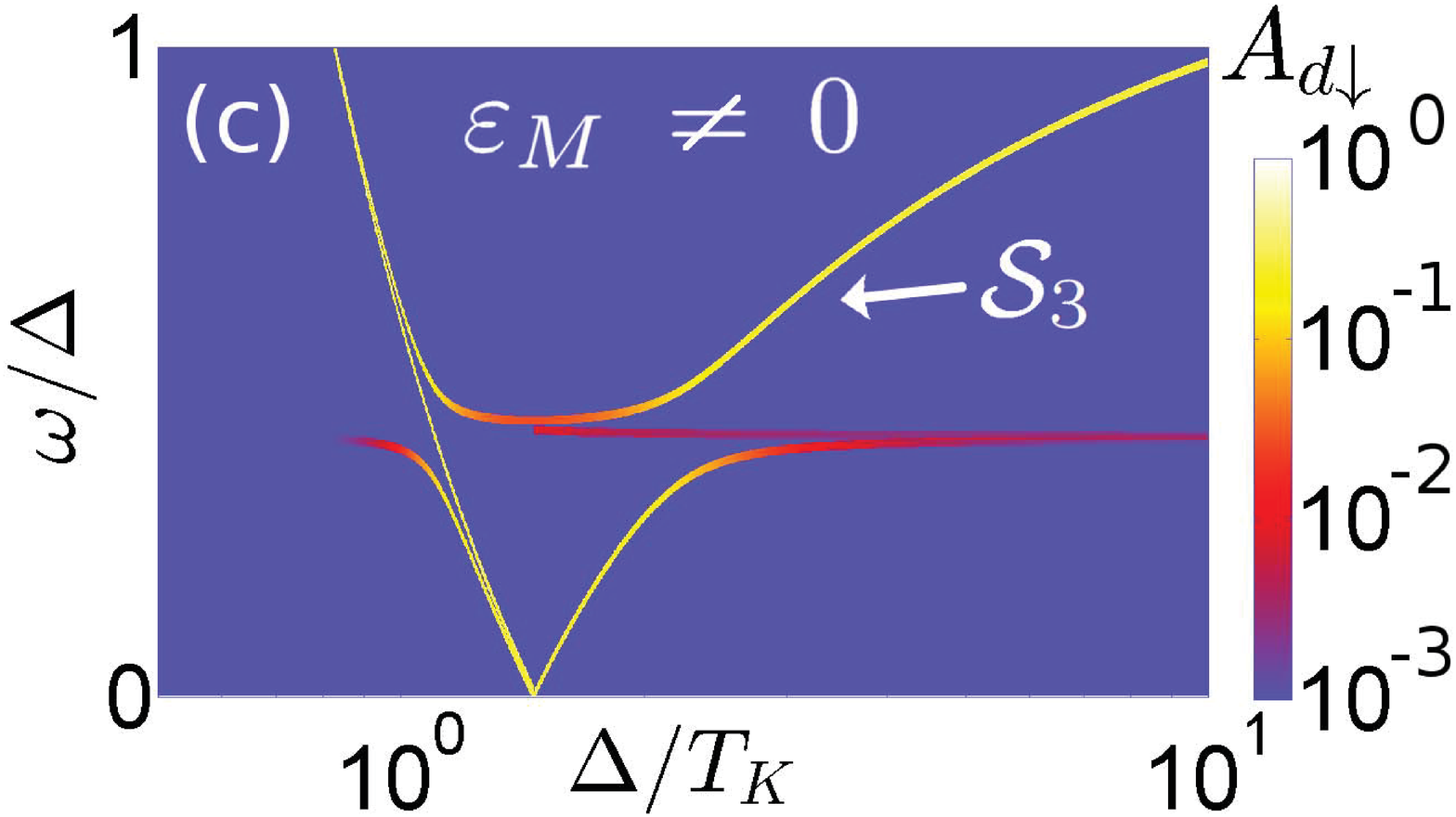}
\includegraphics[width=0.49\columnwidth]{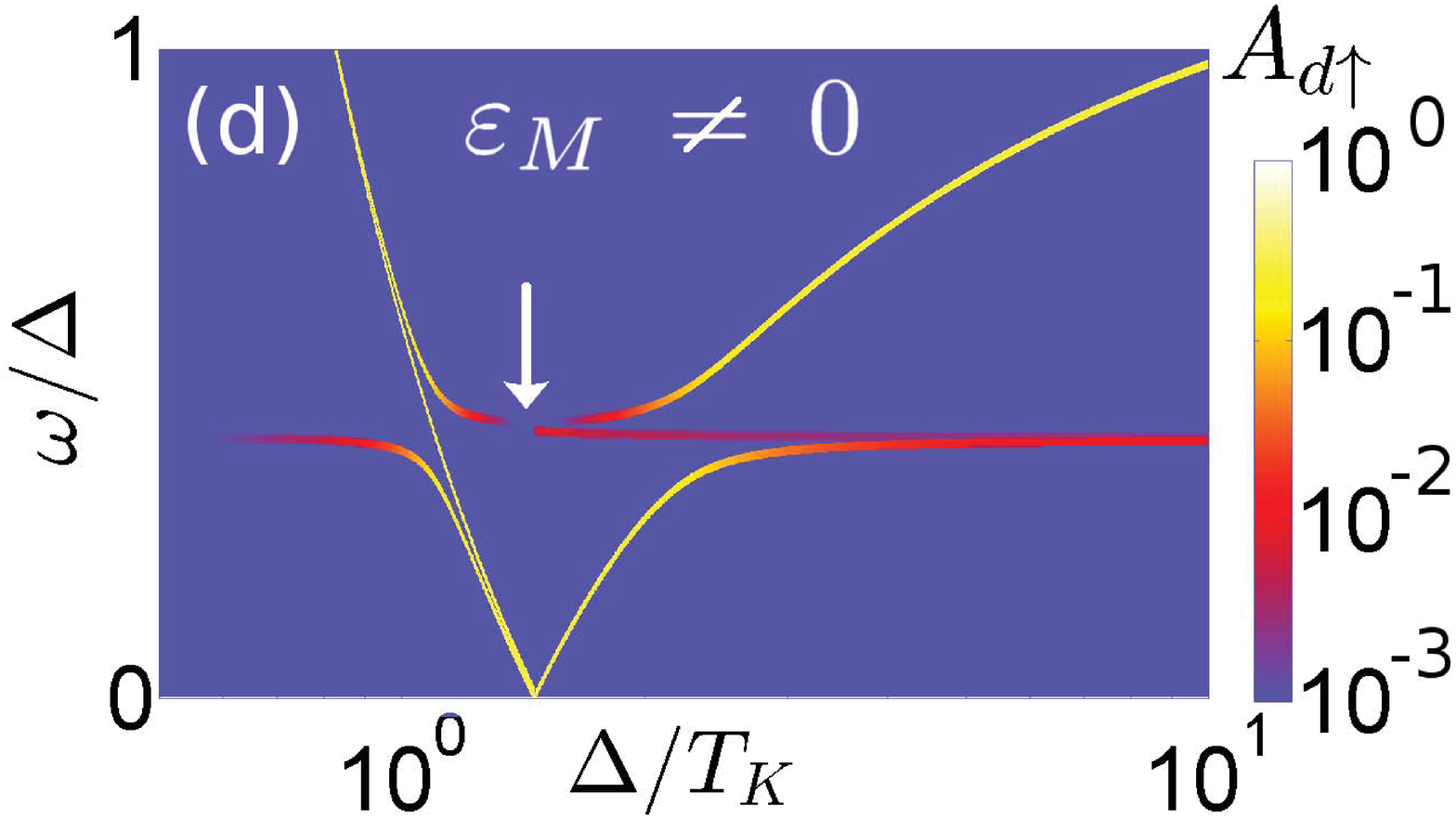}
\includegraphics[width=0.96\columnwidth]{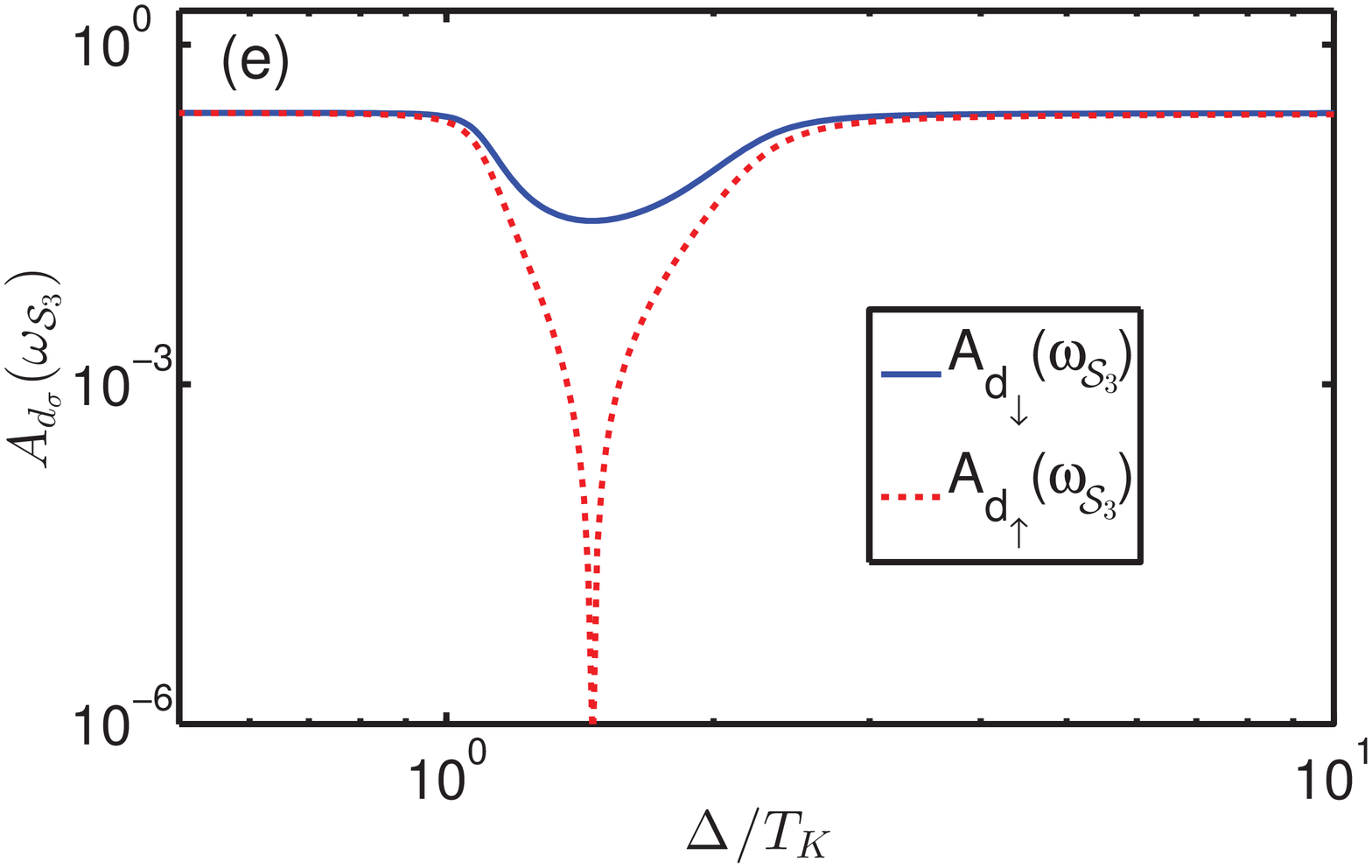}
\caption{(Color online) Panels (a-b): Evolution of the spectral functions  
$A_{d_{\downarrow}}(\omega)$ and $A_{d_\uparrow}(\omega)$ in the subgap region
$|\omega|<\Delta$ for $\varepsilon_{M}=0$. 
(c-d) The same as in (a-b) but for finite $\varepsilon_{M}=0.4\;\Delta$, represented only in the positive frequency domain $0<\omega<\Delta$. The spectral functions present 
electron-hole symmetry, and the spectra for the positive and negative
frequencies are symmetrical. 
(e) Evolution of the peak 
weights for $A_{d_{\sigma}}(\omega_{\cS_{3}})$ as we follow the $\cS_{3}$ resonance indicated in panel (c).
At the QPT, the spin-$\uparrow$ weight vanishes. The white arrows in panels (b) and (d)
indicate where the spin imbalance takes place.
In all the panels, the results are analytical estimates in the atomic limit (see Appendix~\ref{app:diagon} for details). Similar results (not displayed here) are obtained with the NRG approach.
} 
\label{fig:At_sp_fun}
\end{figure}

When $V_{M}\ne 0$ but $\varepsilon_{M}=0$, the global $SU(2)$ spin symmetry is completely lost \footnote{The Majorana mode $\gamma_{1}$ is 
coupled to the spin-down channel only, so that the $U(1)$ symmetry corresponding to the conservation of $S_{z}$ is also broken.} and the states can be organized by parity only (see Appendix \ref{app:diagon} for explicit expressions). In what follows, we shall  label  $\cE_{i} (\cO_{i})$, with $i=g, 1,2, \dots$ the states in the even (odd) sectors. If $U$ is sufficiently large compared to $\Delta$, the highest two energy levels will be in the continuum and only six states lie inside the gap (see Fig.~\ref{fig:levels_at}(b)). They come in pairs such that each even state is degenerate in energy with an odd state. Since the ground state has no defined parity, the QPT is completely washed away. In this limit, the spectral function of the $d^{\dagger}_{\uparrow}$-operator in the QD shows a resonance pinned at $
\omega=0$ which corresponds to the $\cE_{g}\leftrightarrow \cO_{g}$ transition (see Fig \ref{fig:At_sp_fun}(b)). We expect this transition to be visible in the differential conductance across the 
dot in the whole $\Delta/T_{K}$ domain~\cite{Deacon.10}. Keeping $\Gamma_{N}$ 
small enough, we rule out any possible Kondo correlations that would give a similar signal.

In general, we can assume that the Majorana fermion may have a finite energy $\e_M$, which is 
related to the physical length $L$ and the coherence length $\zeta$ of the TSW, and scales as $\e_M\propto  E\exp(-L/\zeta)$ \cite{Alicea.12, Dumitrescu.15}, where $E$ is given by the product of the induced gap and the momentum at the Fermi level \cite{Brouwer.11}. In current experiments, $\e_M$ is likely in the range 100-200 mK \cite{Sarma.15}, however it can in principle be adjusted by changing the parameters that affect the coherence length and control the transition to the topological phase, such as the Zeeman splitting or the chemical 
potential. In this case, the states $\cE_{i}$ and $\cO_{i}$ are no longer degenerate in energy and the QPT is restored. When $\Delta\sim T_{K}$, the ground state changes from an even to an odd state. The  energy spectra can again be found analytically. Their evolution is presented in Fig.~\ref{fig:levels_at}(c). We have found that the MBS presents unique signatures when 
compared with other possible configurations (not presented 
here). Thus, when the MBSs are for example replaced by a resonant level with an on-site 
$U=\infty$, the system has a restored $U(1)$ symmetry corresponding to the conservation of 
$S_{z}$, and the energy spectrum is affected significantly. For the sake of completeness, we present in Fig.~\ref{fig:levels_at}(d) the energy spectrum 
using the same set of parameters as those in panel (c), but computed with the NRG approach. The agreement between the results guarantees the correctness of the superconducting atomic limit.

\section{Spectral Functions}\label{sec:sp_fun}

The energy spectrum discussed so far can be captured in subgap spectroscopy when the 
normal lead is weakly coupled to the dot, i.e., $\Gamma_{N}\ll \Gamma_{S}$.
Recently, signatures of the MBSs formed 
inside a vortex core in a $\rm Bi_{2}Te_{3}/Nb Se_{2}$
heterostructure have been detected by using spin-polarized scanning tunneling 
spectroscopy (SP-STS)~\cite{Sun.16}. They have been  revealed
by measuring the differences in the differential conductance
with the tip magnetization aligned along/against the local magnetic field. 
In our setup, the coupling between the TSW and the QD is spin-selective, as only the 
spin-$\downarrow$ channel is coupled to the MBSs.  Then, we expect a similar imbalance in the 
spin resolved spectral functions. This imbalance can be observed by performing spin-polarized transport measurements by replacing the normal lead with a ferromagnetic one in the setup presented in Fig.~\ref{fig:sketch}.

In this section we discuss the results for the 
spin-polarized spectral functions $A_{d_{\sigma}}(\omega)$ of the operators
$d^{\dagger}_{\sigma}$ describing the excitations in the dot. They are defined as
\begin{equation}
A_{d_{\sigma}}(\omega)=-{1\over \pi} \rm{Im}\, G_{d_{\sigma}}(\omega)\,,
\end{equation}
with $G_{d_{\sigma}}(\omega)$ being the Fourier transform of the retarded electronic Green's function: 
$G_{d_{\sigma}}(t) = -i\theta(t)\langle\{d_{\sigma}(t), d^{\dagger}_{\sigma}\} \rangle$.
We present results  for zero temperature, in which case 
$A_{d_{\sigma}}(\omega)$ captures transitions between the ground state and 
the excited Shiba states having different parities. Moreover, we are 
interested in distinctive characteristics due to the coupling to the MBS, so that we discuss only the
subgap spectrum $\omega<\Delta$.  
In Fig.~\ref{fig:At_sp_fun}(a,b) and Fig.~\ref{fig:At_sp_fun}(c,d) we represent the spin resolved spectral functions $A_{d_{\sigma}}(\omega)$ for $\varepsilon_{M}=0$ and for finite $\varepsilon_{M}$ respectively. All the results presented in Fig.~\ref{fig:At_sp_fun}
were obtained analytically within the atomic limit. 
For the sake of completeness, we have also performed NRG calculations (not displayed here) that confirm 
the main features. The expressions for the energy eigenstates
and the matrix elements of the $d_{\sigma}$ operator are given in Appendix~\ref{app:diagon}. The Shiba states are mixed with the MBSs,
and together they form the localized states inside the gap. Therefore,
they contribute with similar weights to the transitions captured by the 
spectral functions. Moreover, the excited states with the same symmetry lead
to the formation of avoided crossings on either side of the QPT. 
The spectral functions present two distinctive features: ({\it i})  For $\varepsilon_{M}=0$, the spectral function $A_{d_{\uparrow}}(\omega)$ always shows a 
resonance at $\omega=0$, which is due to the $\cE_{g}\leftrightarrow \cO_{g}$ transition, 
while $A_{d_{\downarrow}}(\omega)$ does so only asymptotically in the large $\Delta/T_K$ limit (in the non-symmetric case, $\e\neq-U/2$, both spectral functions have a 
resonance at $\omega=0$);
({\it ii}) For both finite and vanishing $\varepsilon_{M}$, there is a small window close to the QPT, where the transition between the ground state and the third excited Shiba state, $\cE_{g}\leftrightarrow \cO_{3}$ or $\cO_{g}\leftrightarrow \cE_{3}$, becomes strongly spin-polarized. This region is indicated by the arrow in Fig.~~\ref{fig:At_sp_fun}(b, d).
Following this resonance, the weight of the 
spectral function for the spin-$\uparrow$ channel vanishes close to the QPT. This is shown in 
Fig.~\ref{fig:At_sp_fun}(e), where we represent the weighs for $A_{d_{\sigma}}
(\omega_{\cS_{3}})$ at the resonance frequency as we follow the $\cO_{3}$ and $\cS_{3}$ resonance.

We want to highlight the fact that this behavior is a clear signature of the Majorana fermions, and is due to the existence of the anomalous hopping term in Eq.~\eqref{eq:HDM} that simultaneously creates (annihilates) two quasiparticles, one on the dot and one on the TSW. Moreover, no such behavior is expected when the QD is side-coupled to a resonant level or to a second QD, for example. 

\section{Concluding Remarks}

A quantum dot coupled to a normal superconductor shows characteristic resonant features in 
the subgap spectrum. These are known as Shiba bound states and have been measured
in tunneling spectroscopy experiments~\cite{Deacon.10}. Such a system has two distinct 
phases separated by a quantum phase transition. 

Majorana fermions are particles that are their own antiparticles, and have been predicted 
in condensed matter systems~\cite{Alicea.12}, and in particular as localized states
at the ends of a topological wire. Their signature is associated with the existence of a zero energy mode, and has been confirmed experimentally in various experiments~\cite{ Nadj-Perge.14, Xu.15, Mourik.12}. 

In the present work we propose a setup in which clear fingerprints of the Majorana
modes can be detected by using spin-resolved tunneling spectroscopy measurement. We have
investigated the changes in the subgap spectrum when the quantum dot 
is side-coupled to such a topological wire. We have found distinctive hallmarks that
can be associated only with Majorana bound states. 
In particular, we have found that when the two Majorana modes at the ends of 
the topological wire are completely decoupled, the quantum phase transition is washed away, but a finite coupling between them restores the phase transition. Moreover, in the vicinity of the quantum phase transition, the spin resolved spectrum for the dot operators becomes strongly spin-polarized. 
\begin{acknowledgments}
This work was supported by the Romanian National Authority for Scientific Research and
Innovation, UEFISCDI, project number PN-II-RU-TE-2014-4-0432, and by the Hungarian research fund OTKA under grant No. K105149.
\end{acknowledgments}
\appendix

\section{Analytical results in the atomic limit}\label{app:diagon}
In this appendix we present analytical details for the calculation of the 
energy spectrum and for the transition amplitudes of the $d_{\s}$-operators between these
states. These quantities are needed for the 
calculation of the spectral functions, which are discussed in Sec.~\ref{sec:sp_fun}. 
In the atomic limit, the Hamiltonian \eqref{eq:H_at_lim} can be diagonalized 
exactly. Altogether, there are eight states, but only six of them reside inside the gap,
the other two merging with the continuum. 
We shall consider the electron-hole symmetrical case, i.e, $\e=-U/2$, for which simple analytical expressions can be found. We shall focus here on the region
$\Delta/T_{K}\gg 1$, where the ground state resides in the odd sector. 
The corresponding six eigenenergies are
\begin{eqnarray}
E^{{\cal O}}_{i=g,1,2}&=&\frac{1}{4}\left(U+2\eta_{i1}\Gamma_{\varphi}+2 \e_M + \right .\nonumber\\
 & &\left .+  \eta_{i2}\sqrt{8V_M^2+(U+2\eta_{i1}\Gamma_{\varphi}+2\e_M)^2}\right)\nonumber\,,\\
E^{{\cal E}}_{i=1,2,3}&=&\frac{1}{4}\left(U+2\eta_{i2}\Gamma_{\varphi}+2 \e_M + \right .\nonumber\\
 & &\left . + \eta_{i3}\sqrt{8V_M^2+(U+2\eta_{i2}\Gamma_{\varphi}-2\e_M)^2}\right)\nonumber\,,
\end{eqnarray}
where the upper superscripts $\{\cal E, \cal O \}$ label the even and respectively odd 
states, and we introduced the notation $\eta_{ij}=-1+2\delta_{ij}$, with $\delta_{ij}$ the 
Kroenecker symbol. Depending on the ratio $\Delta/T_{K}$, the ground state changes from an
even to an odd state across the QPT point. The energies have to be rescaled so that the ground state has zero energy. 

At the same time, we can obtain the eigenstates exactly. Within the basis $\{$ $\ket {0}, \ket{\uparrow},\ket{\downarrow},\ket{\uparrow \downarrow} \}\otimes \{\ket{0_M}, \ket{1_M}$ $\}$, formed out of the dot and Majorana states, the eigenvectors are given by
\begin{eqnarray}
\ket {{\cal E}_{i=1,2,3}}= \frac{1}{\sqrt{2(u_{{\cal E}_{i}}^2+1)}} &&\left ( u_{{\cal E}_{i}}\left(\ket {0}-\eta_{i2}\ket{\uparrow \downarrow}\right )\ket{0_M}+\right .\nonumber \\
&&\left . +\left(-\eta_{i2}\ket {\uparrow}+\ket{\downarrow}\right)\ket{1_M}\right )\nonumber\,,\\
\ket {{\cal O}_{i=g,1,2}}= \frac{1}{\sqrt{2(u_{{\cal O}_{i}}^2+1)}} &&\left ( u_{{\cal O}_{i}}\left(\ket {\uparrow}-\eta_{i1}\ket{\downarrow}\right )\ket{0_M}+\right .\nonumber \\
&&\left . +\left(-\eta_{i1}\ket {0}+\ket{\uparrow \downarrow}\right)\ket{1_M}\right )\nonumber\,,
\end{eqnarray}
where we used the short hand notations
\begin{eqnarray}
u_{{\cal E}_{i=1,2,3}}=&&\frac{1}{2\sqrt{2}V_M}\left( U+2\eta_{i2}\Gamma_{\varphi}-2\e_M +
\right .\nonumber\\
&&\left . +\eta_{i3}\sqrt{8V_M^2+(U+2\eta_{i2}\Gamma_{\varphi}-2\e_M)^2}\right )\,,\nonumber\\
u_{{\cal O}_{i=g,1,2}}=&& 2\sqrt{2}V_M/\left(U+2\eta_{i1}\Gamma_{\varphi}+2\e_M + \right .\nonumber \\
&&\left .+\eta_{i2}\sqrt{8V_M^2+(U+2\eta_{i1}\Gamma_{\varphi}+2\e_M)^2} \right )\nonumber\,.
\end{eqnarray}
This allows us to compute the matrix elements for the operators in the dot. Since 
$d_{\sigma}$ is a charge Q=1 operator, only transitions between states with 
different symmetry are allowed. Furthermore, since the Majorana mode couples only to the 
spin-$\downarrow$ channel, the matrix elements for $d_{\uparrow}$ and $d_{\downarrow}$ 
are different. For example: $\bra {{\cal{O}}_g}d_{\uparrow} \ket{{\cal{E}}_3}
=(1+u_{{\cal{E}}_3}u_{{\cal{O}}_g})/2\sqrt{(u_{{\cal{E}}_3}^2+1)(u_{{\cal{O}}_g}^2+1)}$ and 
$\bra {{\cal{O}}_g}d_{\downarrow} \ket{{\cal{E}}_3}=(1-u_{{\cal{E}}_3}u_{{\cal{O}}_g})/
2\sqrt{(u_{{\cal{E}}_3}^2+1)(u_{{\cal{O}}_g}^2+1)}$. Then, the spectral functions are
simply given by 
\begin{equation}
A_{d_{\sigma}}(\omega) = |\bra {{\cal{O}}_g}d_{\s} \ket{{\cal{E}}_3}|^{2}\delta(\omega +E_{g}^{\cal O}-E_{3}^{\cal E}). 
\end{equation}
Their evolution across the QPT is represented in Fig.~\ref{fig:At_sp_fun}(e). Notice that the spin-up spectral function becomes zero at the QPT point corresponding to $U=2\Gamma_{\varphi}$.

\bibliography{references}

\end{document}